# Chirality meets Topology

**By considering the topology of chiral crystals a new type of massless fermions, connected with giant arc-like surface states, are predicted. Such Kramers-Weyl fermions should manifest in a wide variety of chiral materials.**

*Chandra Shekhar- Max Planck Institute for Chemical Physics of Solids, Germany*
*shekhar@cpfs.mpg.de*

Two objects, which are mirror images and cannot be superimposed on each other, are chiral objects. The most familiar example is, perhaps, our own hands: left and right. The concept of chirality has a long history dating back to the ancient Greeks: indeed the Greek word *cheir*, means 'handedness'. Another familiar example is spirals or helixes which can twist one way or the other. For example the DNA molecule can take only one of two possible chiral structures (see Fig. 1 (a)) in all living organisms [1]. At the atomic level, crystals are formed from periodic arrays of unit cells that have distinct symmetries such as inversion, mirror or roto-inversion: crystals that lack these symmetry elements may be chiral. Fig 1(b) shows one example of a chiral crystal, namely, $Ag_2Se$. Combining this concept of chirality and modern theories of the topology of condensed matter, Guoqing Chang and colleagues [2] computed the topological electronic properties of a large number of nonmagnetic chiral compounds such, as $Ag_2Se$, and found that they host a new type of fermion, termed Kramers-Weyl fermions.

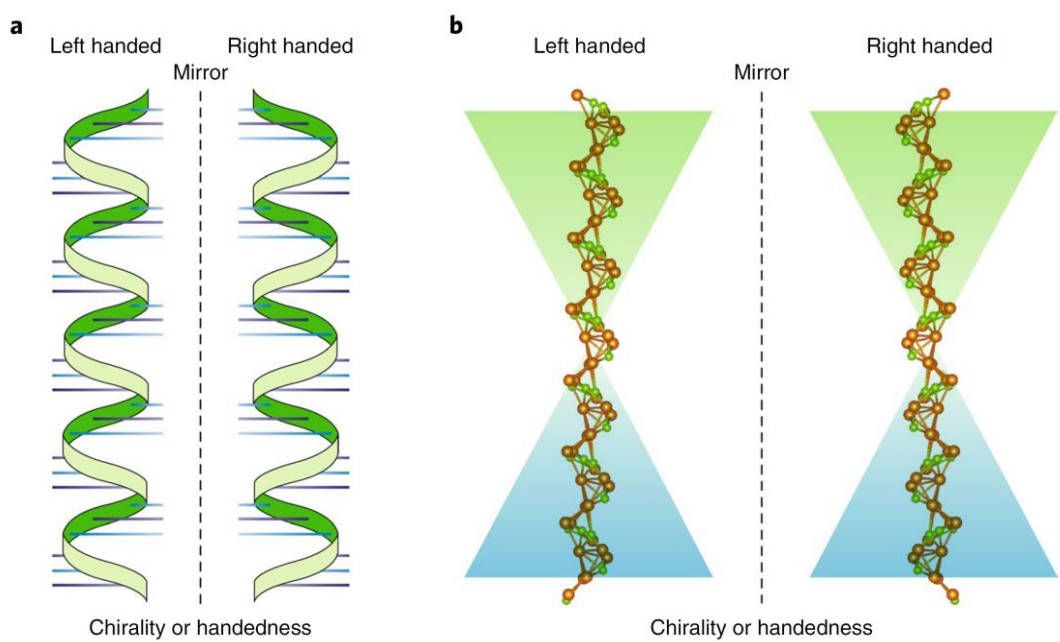

**Fig. 1** | **Examples of chiral structures**. **a**, Free-standing molecules of DNA. The lines represent the DNA bases AGTC. **b**, Crystal structure of the chiral material $Ag_2Se$ and Dirac

cone — a fingerprint of massless fermions.

A rich avenue of inquiry in modern condensed matter physics is that of the role of topology of electronic states in influencing or even controlling the properties of a remarkably large set of materials. Topology is determined not only by crystal structure symmetry but also by the symmetry and distribution of the atomic orbitals inside a crystalline material [3]. Using advanced theoretical techniques and powerful computers, thousands of materials have been predicted to be topological by analyzing approximately 500,000 materials in readily available databases. One of the major breakthroughs using the concept of topology has been the identification of massless particles. Several distinct flavors have been identified and already discovered: these are the Weyl, Dirac and Majorana fermions. Among these fermions, Weyl fermions are chiral and indeed always appear in pairs, each with an opposite chirality - much like a source or sink of the magnetic field of a magnetic monopole. In condensed matter physics, the enormous range of known and still unknown materials with various symmetries and topologies have spawned not only the novel fermions above but also even more that have yet to be found [4].

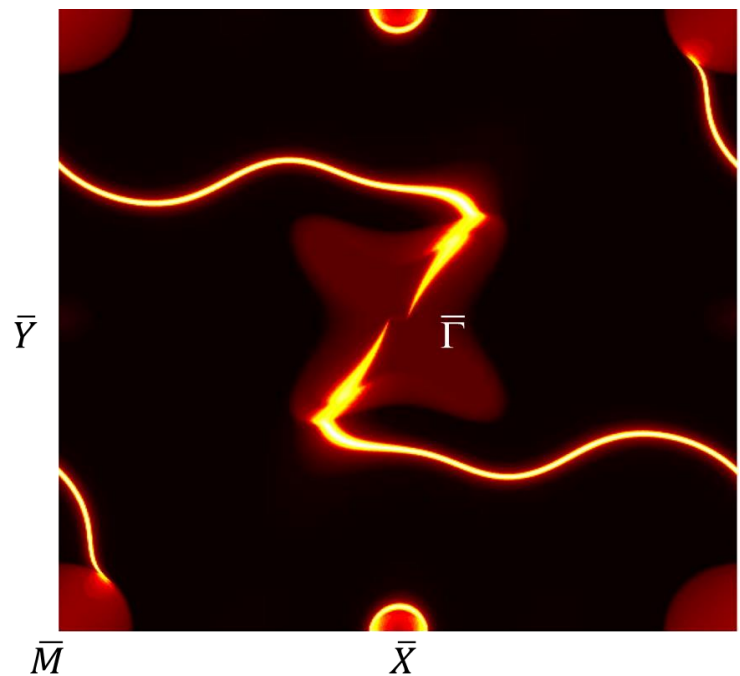

**Fig. 2 | Kramers–Weyl fermions.** Gigantic Fermi arcs (bright yellow lines) connecting boundaries of Brillouin zone of a CoSi chiral crystal

Indeed by explicitly considering chiral and topological effects in crystalline materials Chang et al have predicted a new Kramers-Weyl class of fermions. In given conditions, these particles appear only at high symmetric points of the Brillouin zone (BZ) - Kramers points. They are protected by symmetry and are connected by multiple arc-like Fermi states, which span a large amount of reciprocal space and exist over a broad energy range as shown in Fig. 2 for CoSi - a member of one of the chiral space groups predicted to host Kramers-Weyl fermions in this study. The surfaces of chiral crystals are highly helicoid that can exhibit more exotic phenomena for example: quantized circular photogalvanic effect. Chang et al. applied the approach of topological quantum chemistry in all 65 chiral space groups (out of 230 space groups) and revealed Kramers-Weyl fermions. Unlike usual Weyl fermions, structural chirality and an internal capacity of atoms (known as spin orbit coupling (SOC)) play a primary role in stabilizing Kramers-Weyl fermions while band inversion is found not to be necessary. Depending on the symmetry of symmorphic chiral space group as well as nonsymmorphic chiral space group, three-fold, four-fold and six-fold crossings can be found with higher chern number, that the authors attribute to higher order Kramers-Weyl fermions [2]. Chern number is directly related to number of arcs connecting Kramers-Weyl points. Such higher order fermions are restricted to exist by Lorentz invariance in high-energy physics while these are allowed by symmetry in crystalline materials. The triple-fold fermions are even very close to observation [5]. Hundreds of known nonmagnetic materials exist in the 65 chiral space groups and some of them are listed by Chang *et al* and this number perhaps will increase dramatically, if magnetisms include in chirality. Now are the tasks of experimentalists to observe the Kramers-Weyl points and their linked arc-like surface sates employing the advance spectroscopic tool (ARPES, angle- and spin-resolved photo emission spectroscopy) and the advance microscopic tool (STM, scanning tunneling microscopy). To accomplish the results, a right material selection is really important because the Kramer-Weyl points and arcs depend on the energy band splitting by SOC and, sometimes, are buried in other spurious Fermi pockets. Indeed, some other promising experiments for example, quantized circular photogalvanic effect and anomalously large spin Hall effect and chiral magnetic effect, are rather indirect but can underline the physics of Kramer-Weyl fermions. Among the plethora of nonmagnetic chiral crystalline materials, a surprising number of simple binary materials have clean electronic structures, and are readily available for example: transition metal monosilicides, AlPd and AlPt, and thus it is anticipated that experimental observations of Kramer-Weyl fermions will take place soon.